\documentclass[prl,twocolumn, showpacs]{revtex4}
\usepackage{graphicx}
\usepackage{graphics}
\usepackage{amssymb}
\usepackage{epstopdf}
\usepackage{color}
\usepackage{subfigure}
\usepackage{epsfig}

\begin{document}
\title{A Steady State Thermodynamic Entanglement Witness}
\author{Jenny Hide}
\affiliation{Pevensey 2 Building, University of Sussex, Falmer Campus, Brighton BN1 9QF, UK \\
Department of Physics, Graduate School of Science, University of Tokyo, Tokyo 113-0033, Japan}

\date{\today}

\begin{abstract}
We present a method for entanglement detection in steady
state systems using a thermodynamic witness. To illustrate, we consider
an example and find that for an $XX$ spin chain, the presence of an
energy current increases the region of entanglement
detected by the steady state witness. Further, we find that entanglement
exists even at a high steady state temperature. We also discuss
concurrence in the steady state system and find that the
\emph{amount} of entanglement can be increased by the current.
\end{abstract}

\maketitle


Entanglement is one of the most fascinating aspects of quantum
physics and is an important resource in the field of quantum
information. While a separable state can be written as a convex sum
of pure product states, $\sigma=\sum_i p_i \sigma_i^1 \otimes
\sigma_i^2 \otimes \cdots \otimes \sigma_i^n$, an entangled state
cannot. Many-body entanglement \cite{manybody} is not easy to
quantify; while thermal pairwise entanglement can be measured using
concurrence, measuring multipartite entanglement is a difficult
task. Entanglement \emph{witnesses} are a useful alternative to
these measures. An entanglement witness is an expectation value of
an operator for which a bound can be found for any separable state.

The subject of quantum thermodynamics \cite{Gemmer} has already
contributed to the success of entanglement witnesses \cite{ent_wit,vlat}.
For example, magnetic susceptibility can witness entanglement
\cite{Wiesniak}, as can an average over the nonequilibrium work done
during a process \cite{HideJ}.
At present, thermodynamic witnesses are used to detect
entanglement in equilibrium systems. Our aim
is to extend the success of equilibrium entanglement witnesses
to nonequilibrium steady state systems.

In this letter, we demonstrate how a steady state entanglement witness can
be found. Specifically, we find a steady state thermodynamic quantity,
a current, which can detect entanglement in steady state systems.
We consider an $XX$ spin chain in the thermodynamic limit
to exemplify this witness and find that
the introduction of an energy current increases the region of entanglement detected.
In order to gain insight into whether the entanglement itself
increases with the driving field, we also calculate the chain's correlation
functions and use them to determine the nearest and
next nearest neighbour concurrence. We discover that the concurrence can
be increased by the presence of the energy current.


Thermodynamics has been used in many areas of physics. It is
limited to equilibrium systems, although some thermodynamic concepts such
as work done apply during dynamical processes.
No equivalent formulation for nonequilibrium thermodynamics exists at
present. However, attempts have been made to provide a framework for steady state
thermodynamics  \cite{Oono,Sasa} since these systems are time
independent and simpler to investigate than the general out
of equilibrium case. There are also methods by which
steady state systems can be modeled.


In particular, the Hamiltonian of an equilibrium system can be modified to allow a steady
state current passing through the system to be described
\cite{Hardy, Saito, Michel}. In this way,
nonequilibrium steady state systems can be mapped onto equilibrium systems.
For an energy current, this is achieved by defining the operator of the current using
continuity equations,
$\partial h(x,t) /\partial t + \partial j(x,t) /\partial x = 0$ where $h(x,t)$
is the energy density operator and $j(x,t)$ is the heat flux operator.
For a chain with $N$ sites, the discrete form of this is $dh_l/dt = j_{l-1}
-j_l$, and the left hand side can be written $dh_l/dt = i[H,h_l]$.
For one dimensional systems $H_0=\sum_{l=1}^N (h_l^0
+ V(l,l+1))$, this has been shown to be \cite{WuAndSegal}

\begin{eqnarray}
j_l  & = & \frac{i}{2} \left( \left[ h_l^0 - h_{l+1}^0, V(l,l+1) \right]
+ \left[ V(l,l+1), \right. \right. \nonumber \\
& & \left. \left. V(l+1,l+2) \right] + \left[ V(l-1,l), V(l,l+1) \right] \right).
\label{eq:GeneralEnC}
\end{eqnarray}

Here, $h_l^0$ is the non-interacting part of the Hamiltonian, and
$V(l,l+1)$ is the interaction. The energy current operator is
$J^E=\sum_l j_l$ and can be incorporated into the Hamiltonian formalism using a
Lagrange multiplier, $\lambda$, so that we have $H=H_0-\lambda J^E$.
In the case that $[J^E,H_0]=0$, and $H_0$ can be diagonalised
analytically, the total Hamiltonian $H$ can also be diagonalised
using the same method.

We note that the origin of the steady state current is not important
when using this method. That is, the current is given by Eq.
\ref{eq:GeneralEnC} whether it is induced by an external driving field or
by a reservoir \cite{Kosov}. For example, the steady state could be achieved
by holding opposite ends of an open spin chain at different, constant,
temperatures, but the formalism above is not limited to this scenario.

Using Eq. \ref{eq:GeneralEnC}, the ground state correlation functions of an
Ising spin chain have been investigated \cite{Antal1,Antal2}.
Also at zero temperature, a steady state $XX$
spin chain with an energy current increases the entanglement
present \cite{Eisler} compared to the equilibrium case,
and quantum state transfer can be improved  \cite{ZWang}.
In this letter, we take this formulation of steady state systems into the thermal
regime using the method outlined below.


It is possible to construct a steady state thermal
density matrix  \cite{Casas, Kita} which is similar in form to the equilibrium case:
$\rho = e^{-\beta H_0 - \gamma J^E}/Z$,
where $Z=\mathrm{tr}(e^{-\beta H_0 - \gamma J^E})$ is the steady state partition
function, $J^E$ is the steady state energy current operator discussed above,
and $\beta$ and $\gamma$ are Lagrange multipliers. $\beta=1/T$ can be
thought of as a generalised inverse temperature which is valid
in steady state systems, while $\gamma = -\lambda \beta$ is the driving
term of the energy current and has no analogue in equilibrium
thermodynamics.

This definition of the steady state density matrix allows a
nonequilibrium steady state entanglement witness of the form

\begin{equation}
W_{ss}= \zeta \frac{\partial}{\partial \gamma} \ln Z = \eta Q,
\label{Eq:SSWit}
\end{equation}

where $\zeta$ and $\eta$ are constants and $Q=\langle J^E \rangle =
\mathrm{tr}(J^E \rho)$ is the expectation value of the energy current, to be
calculated. Thus we can detect entanglement in the steady state
system using the energy current, a steady state quantity. We note
that other currents, such as a magnetisation current, can be
calculated in a similar way to that described above. Therefore other
currents can be used to witness entanglement similarly to $Q$ in Eq.
\ref{Eq:SSWit}.


To illustrate how the witness works, we consider an $XX$ spin chain in the
thermodynamic limit,

\begin{equation}
H_{XX}= -\frac{J}{2} \sum_l \left( \sigma_l^x \sigma_{l+1}^x + \sigma_l^y
\sigma_{l+1}^y \right) - B\sum_l \sigma_l^z,
\label{Eq:HXX}
\end{equation}

where $J$ is the coupling strength between nearest neighbour sites
and $B$ is an external magnetic field. Using Eq. \ref{eq:GeneralEnC},
the energy current for $H_{XX}$ is

\begin{eqnarray}
J^E & = & -BJ\sum_l \left( \sigma_l^y \sigma_{l+1}^x - \sigma_l^x
\sigma_{l+1}^y \right) \\ \nonumber
& + & \frac{J^2}{2} \sum_l \left( \sigma_l^y
\sigma_{l+1}^z \sigma_{l+2}^x - \sigma_l^x \sigma_{l+1}^z
\sigma_{l+2}^y \right).
\label{Eq:HJE}
\end{eqnarray}

We consider $J$ and $B$ to be positive throughout the letter. The
total Hamiltonian, $H=H_{XX}+ \frac{\gamma}{\beta} J^E$ can be
diagonalised \cite{Katsura} using a Jordan-Wigner transformation,
$a_l=\prod_{m=1}^{l-1} \sigma_m^z \otimes
\left(\sigma_l^x-i\sigma_l^y \right)/2$ and a Fourier transform,
$a_l^{} =\frac{1}{\sqrt{N}} \sum_k d_k e^{\frac{2 i \pi k l}{N}}$ to
give $H=\sum_k (2\Lambda_k d_k^\dagger d_k -\textbf{1} B)$ where
$\Lambda_k \rightarrow \Lambda(q) = (B-J\cos q)(2
\frac{\gamma}{\beta}J\sin q +1)$ as $N \rightarrow \infty$. Although
we consider the thermodynamic limit, our method for calculating a
steady state witness can be applied to a system of any size.

To aid in our later calculations for the witness, we introduce an extra term,
$b$, into the Hamiltonian such that Eq. \ref{Eq:HXX} becomes
$H_{XX}=-\frac{J}{2} \sum_l \left( \sigma_l^x \sigma_{l+1}^x +
\sigma_l^y \sigma_{l+1}^y \right) - bB\sum_l \sigma_l^z$. The total
Hamiltonian, $H$, can again be diagonalised using the
transformations described above. The partition function is thus $\ln
Z = \frac{N}{2\pi} \int_0^{2\pi} dq \ln \left[ 2 \cosh \left( \beta
\xi (q) \right) \right]$ where $\xi (q) = Bb-J\cos q
+2\frac{\gamma}{\beta} JB\sin q - 2\frac{\gamma}{\beta}J^2 \sin q
\cos q$. Once the calculations are performed, we set $b=1$ in order
to recover the original system.

We now introduce two entanglement witnesses. The first is similar to existing
witnesses, \cite{ent_wit,vlat} and is found by calculating the expectation value
of the Hamiltonian
and rearranging the resulting equation. The second is found using Eq. \ref{Eq:SSWit}
and is a genuine steady state witness. We consider both to demonstrate why
the steady state witness is important.

The first witness is

\begin{equation}
W_1= 2\left| \frac{U+BM-\frac{\gamma}{\beta}Q}{J N}
\right|
\label{Eq:NormWit}
\end{equation}

where $U=\langle H \rangle$ is the steady state equivalent of
internal energy, $U=-\frac{\partial}{\partial \beta} \ln Z
-\frac{\gamma}{\beta}\frac{\partial}{\partial \gamma} \ln Z$. The
second term, $-\frac{\gamma}{\beta}\frac{\partial}{\partial \gamma}
\ln Z$ is equivalent to $\frac{\gamma}{\beta}Q$ and therefore
cancels with the final term in Eq. \ref{Eq:NormWit}. We also have
$-\frac{\partial}{\partial \beta} \ln Z=\frac{N}{2\pi} \int_0^{2\pi}
dq (J\cos q - B)\tanh (\beta \Lambda (q))$. We use the convention
that the magnetisation is given by $M= \sum_l \langle \sigma_l^z
\rangle= \left. \frac{1}{\beta B N} \frac{\partial}{\partial b} \ln
Z \right|_{b=1}$ rather than by $\frac{1}{\beta}
\frac{\partial}{\partial B} \ln Z$, though each definition gives the
same witness on appropriate consideration of the remainder of the
expectation value of the Hamiltonian. Thus $M/N =  \int_0^{2\pi}
\frac{dq}{2\pi} \tanh \left[ \beta \Lambda (q) \right]$.

The system is entangled when $W_1 >1$. This bound is calculated
using the expectation value, $| \langle
\sigma_l^x \sigma_{l+1}^x + \sigma_l^y \sigma_{l+1}^y \rangle |$,
which is equal to Eq. \ref{Eq:NormWit}. The
bound is found for pure product states \cite{ent_wit,vlat} using the
Cauchy-Schwarz inequality, $(\sum_l x_l y_l )^2 \leq \sum_l x_l^2
\sum_l y_l^2$ and the definition of a single site density matrix which
leads to the inequality $\langle \sigma_l^x \rangle^2+\langle \sigma_l^y
\rangle^2+ \langle \sigma_l^z \rangle^2 \leq 1$. Since the set of separable states is
convex, this bound is also true for all separable states.

\begin{figure}[t]
\begin{center}
\centerline{
\includegraphics[width=3.3in]{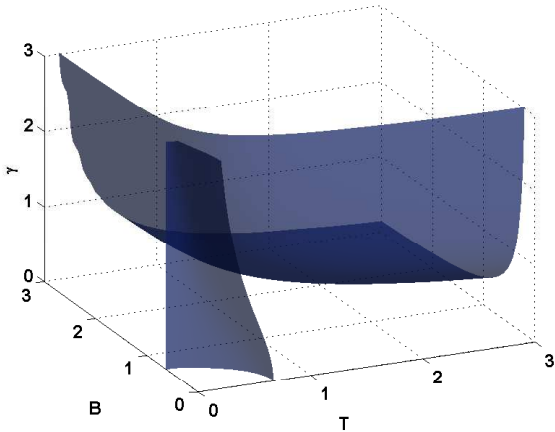} }
\end{center}
\caption{The surface in the plot corresponds to $W_{1}= 1$. The witness detects entanglement
within the convex regions where $W_{1} >1$. We consider $J=1$ and $T$ is
the generalised steady state temperature. }
\label{Fig:EntWitNormal}
\end{figure}

Fig. \ref{Fig:EntWitNormal} shows the region of entanglement
detected by witness $W_1$. It shows that increasing the driving
quantity $\gamma$ increases the region of entanglement detected by
the witness. This is an interesting result since it indicates that
introducing an energy current increases the entangled region in this
thermal system. That is, $\gamma$ increases which values of $B$, $T$
etc. the system is entangled for. In addition, we see that
entanglement can exist even at high steady state temperatures with a
large enough $\gamma$. However, witness $W_1$ is not easily
experimentally measured since $U$ and $M$ are now steady state
thermodynamic quantities and thus no longer correspond to clear
measurable quantities. Therefore, we use Eq. \ref{Eq:SSWit} to
calculate a genuine steady state entanglement witness,

\begin{equation}
W_{ss} = \frac{2 |Q| }{JN(2B+J)}.
\end{equation}

\begin{figure}[t]
\begin{center}
\centerline{
\includegraphics[width=3.3in]{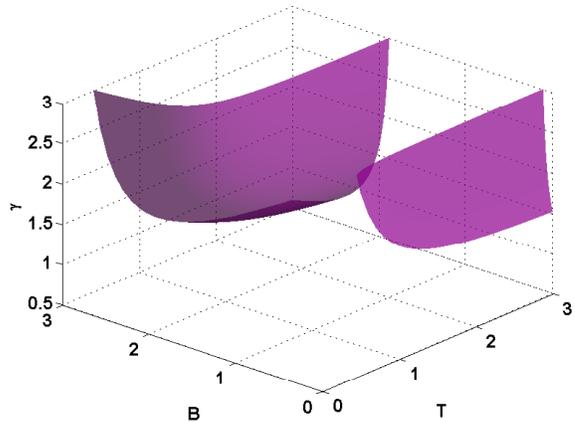} }
\end{center}
\caption{ The surface in the plot corresponds to $W_{ss}= 1$. The witness detects entanglement
within the convex regions where $W_{ss} >1$. We consider $J=1$ and $T$ is
the generalised steady state temperature.}
\label{Fig:EntWitSS}
\end{figure}

Here we use the average energy current, $Q$, itself to
detect entanglement. Since $Q$ is the expectation value of Eq. \ref{Eq:HJE},
we can again use the Cauchy-Schwarz inequality and the definition
of the density matrix to find a bound for separable states.
This leaves $|Q| \leq BJN+\frac{J^2}{2} \sum_l
| \langle \sigma_l^z \rangle |$ for pure product states. For any pure
product state, $| \langle \sigma_l^z \rangle | \leq 1$, hence we can
rearrange the inequality and find that entanglement can be detected when
$W_{ss}>1$. The expectation value of the energy current is
$Q = \frac{NJ}{\pi} \int_0^{2\pi} dq (J\cos q -B) \sin q \tanh (\beta \Lambda (q))$.
The steady state witness is shown in Fig. \ref{Fig:EntWitSS}.
Again increasing $\gamma$ increases the region of entanglement
detected, and entanglement is detected even at high steady state temperatures.
For this witness, no region is detected at low values of $B$ and $T$,
but an extra region exists when $B$ is close to zero and higher $T$.
An interesting way to view this witness is to consider
$W_{ss}$ itself. It shows a large, absolute, expectation value of the
energy current will detect entanglement (after scaling by the constant
$\eta=2/[JN(2B+J)]$).

How the expectation value of the energy current would be measured
experimentally is dependent on the cause of the steady state. For
example, for a spin chain in a constant temperature gradient, the
heat conduction and hence $Q$ can be experimentally determined
\cite{expQ}.


By considering the concurrence, we can confirm that it is the
\emph{amount} of entanglement which increases with $\gamma$.
Concurrence quantifies entanglement between two mixed qubits,
and is given by ${\mathcal C}(\rho) = \max \{0,\lambda_1 -
\lambda_2 - \lambda_3 - \lambda_4 \}$ where $\lambda_1 \geq
\lambda_2 \geq \lambda_3 \geq \lambda_4$.
The $\lambda_i$s are the square roots of the eigenvalues
of the matrix $\rho \tilde{\rho}$ where
$\tilde{\rho} = (\sigma^y \otimes \sigma^y) \rho^* (\sigma^y \otimes
\sigma^y)$.
Since $[H,\sum_l \sigma_l^z]=0$,
the concurrence is given by  $C(\rho_{l,l+R})=2\max\{|z|-\sqrt{vy},0\}$
\cite{OConnor} where
$z=(\langle \sigma_l^x \sigma_{l+R}^x + \sigma_l^y
\sigma_{l+R}^y \rangle -i \langle \sigma_l^y \sigma_{l+R}^x -
\sigma_l^x \sigma_{l+R}^y\rangle )/4$, $vy = ( (1+
\langle \sigma_l^z \sigma_{l+R}^z \rangle)^2 -4\langle \sigma_l^z
\rangle^2)/16$, and we have
used that $\langle \sigma_l^z \rangle=\langle \sigma_{l+1}^z \rangle$ due
to the chain's translational invariance.

We calculate each of the correlation functions following the
method in  \cite{Barouch}, and calculate the nearest neighbour, $R=1$,
and next nearest neighbour, $R=2$, concurrence. Generally,
$\langle \sigma_l^x \sigma_{l+R}^x + \sigma_l^y \sigma_{l+R}^y \rangle
= -\langle A_l \prod_{m=1}^{R-1} (A_{l+m} B_{l+m}) B_{l+R} \rangle +
\langle B_l \prod_{m=1}^{R-1} (A_{l+m} B_{l+m}) A_{l+R} \rangle$,
$\langle \sigma_l^y \sigma_{l+R}^x - \sigma_l^x \sigma_{l+R}^y \rangle
= -i e^{i \pi R} \langle A_l \prod_{m=1}^{R-1} (A_{l+m} B_{l+m}) A_{l+R} \rangle +
i e^{i \pi R} \langle B_l \prod_{m=1}^{R-1} (A_{l+m} B_{l+m}) B_{l+R} \rangle$,
$\langle \sigma_l^z \sigma_{l+R}^z \rangle = \langle A_l B_l A_{l+R}
B_{l+R} \rangle $ and $\langle \sigma_l^z \rangle = \langle A_l B_l \rangle$
where $A_l=(a_l^\dagger +a_l)$ and $B_l=(a_l^\dagger - a_l)$.
Using Wick's theorem,  we can rewrite each of these in terms of two point
correlation functions. For example, the $zz$ correlation function is
$\langle A_l B_l A_{l+R} B_{l+R} \rangle = G_0^2 - G_R^2 +S_R^2$ where
we have defined $G_R= \langle A_l B_{l+R} \rangle$ with
$G_R=- \langle B_l A_{l+R} \rangle$ and $S_R = \langle B_l B_{l+R} \rangle
= -\langle A_l A_{l+R} \rangle$.

Using the same method of diagonalisation as for the Hamiltonian,
we can write $\langle \sigma_l^z \rangle = \frac{1}{N} \sum_k \langle
\textbf{1} - 2 d_k^\dagger d_k \rangle$. In addition, $G_R = \frac{1}{N} \sum_k
\cos (\frac{2\pi k R}{N}) \langle  \textbf{1} - 2 d_k^\dagger d_k \rangle$
and $S_R = \frac{i}{N}  \sum_k \sin (\frac{2\pi k R}{N})  \langle  \textbf{1}
- 2 d_k^\dagger d_k \rangle$. Therefore, the thermodynamic expressions for
$G_R$ and $S_R$
are found directly from the magnetisation, $M/N=\langle \sigma_l^z \rangle$
calculated previously:

\begin{eqnarray}
G_R & = & \int_0^{2\pi} \frac{dq}{2\pi} \cos(qR) \tanh(\beta \Lambda) \\
S_R & = & i\int_0^{2\pi} \frac{dq}{2\pi} \sin(qR) \tanh(\beta \Lambda).
\end{eqnarray}

\begin{figure}[t]
\centering
\begin{tabular}{cc}
\epsfig{file=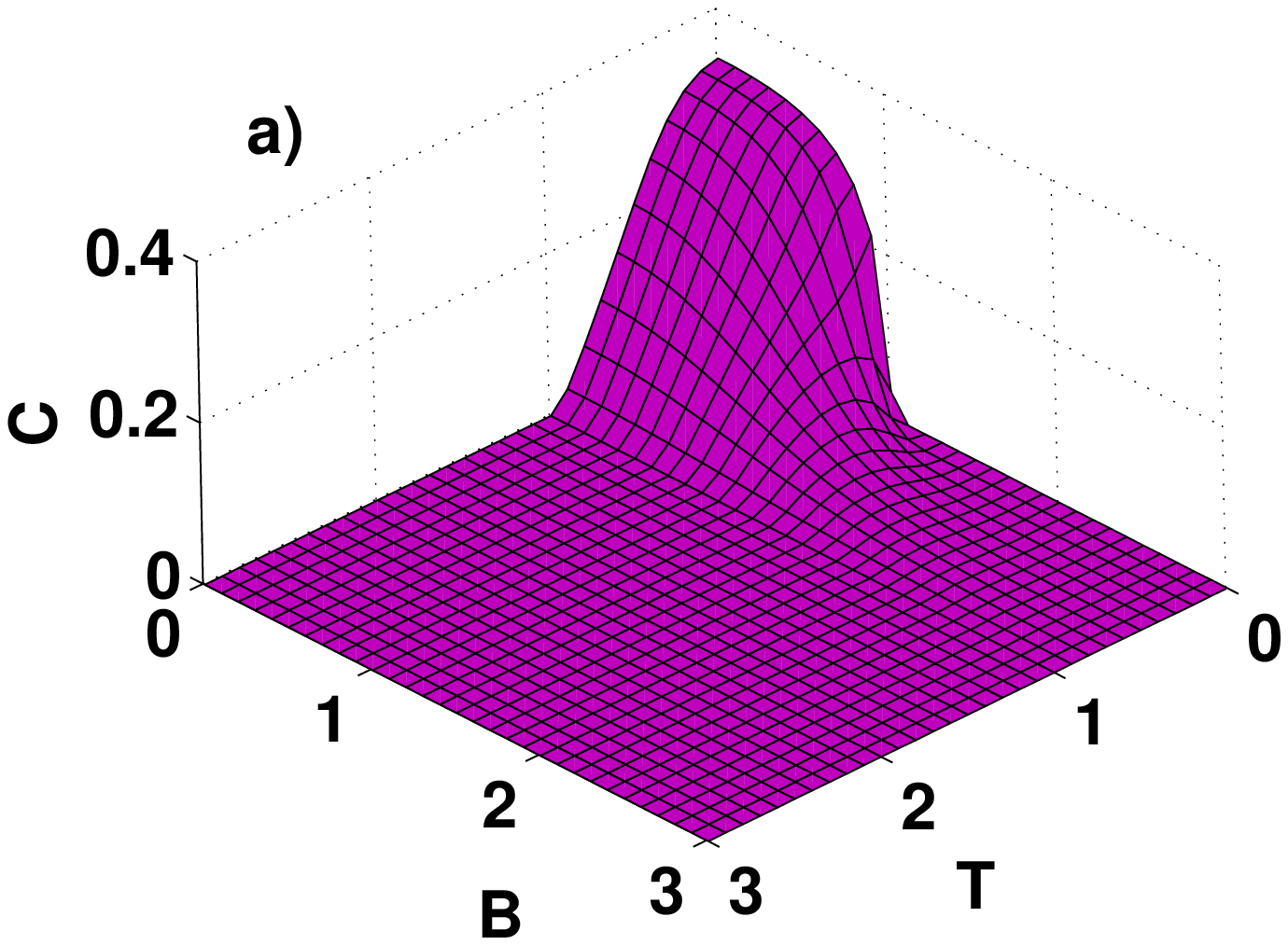,width=0.5\columnwidth,clip=} &
\epsfig{file=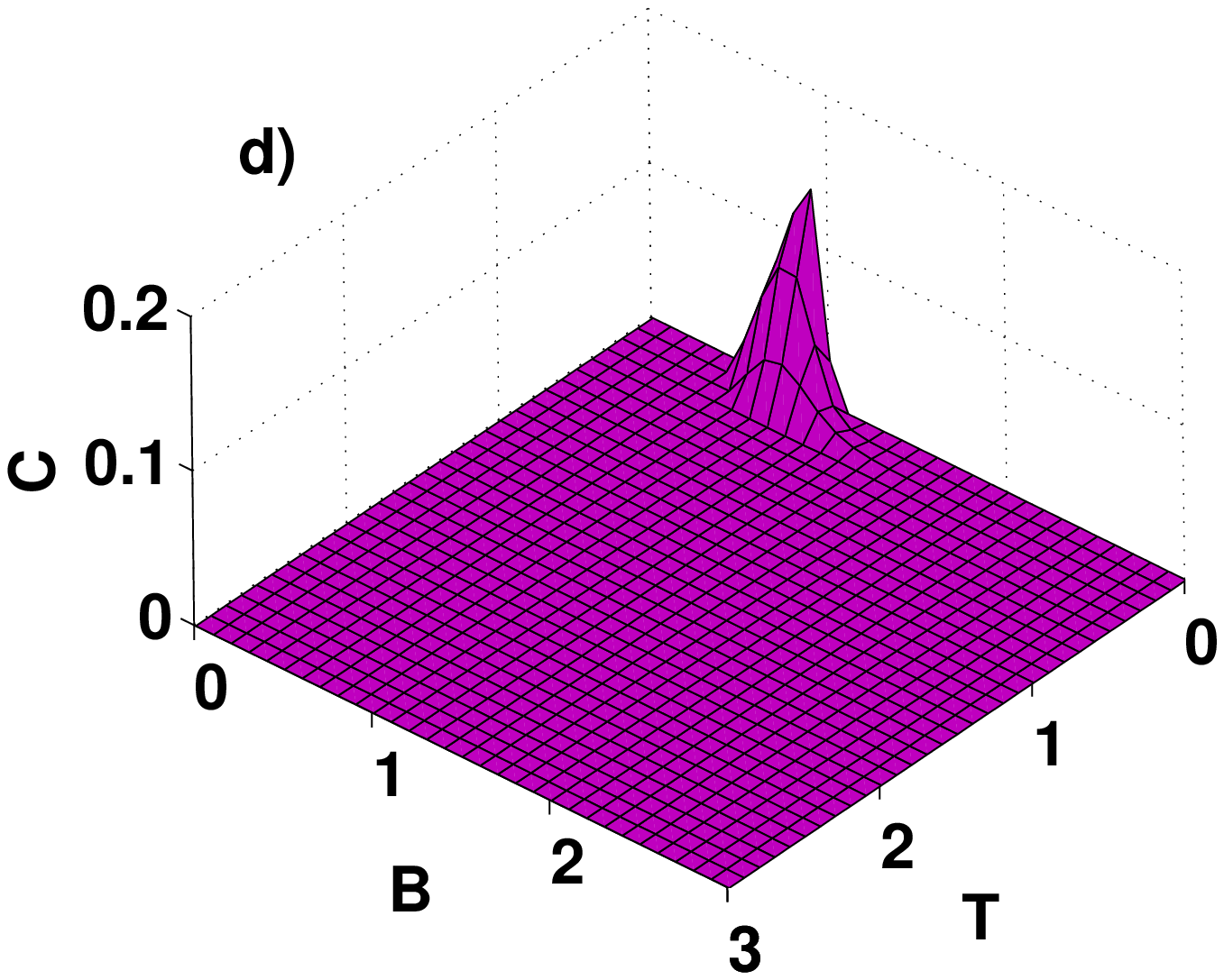,width=0.5\columnwidth,clip=} \\
\epsfig{file=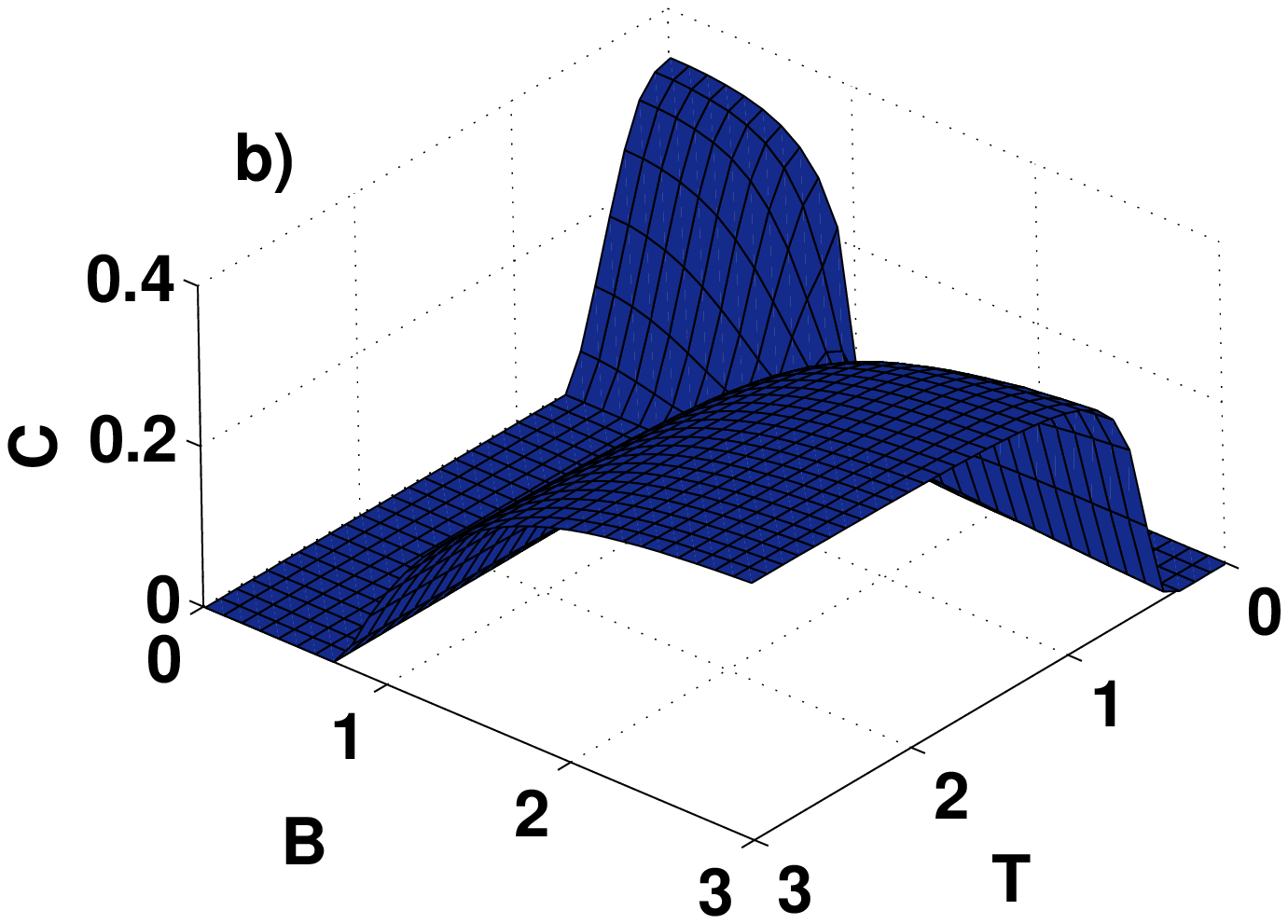,width=0.5\columnwidth,clip=} &
\epsfig{file=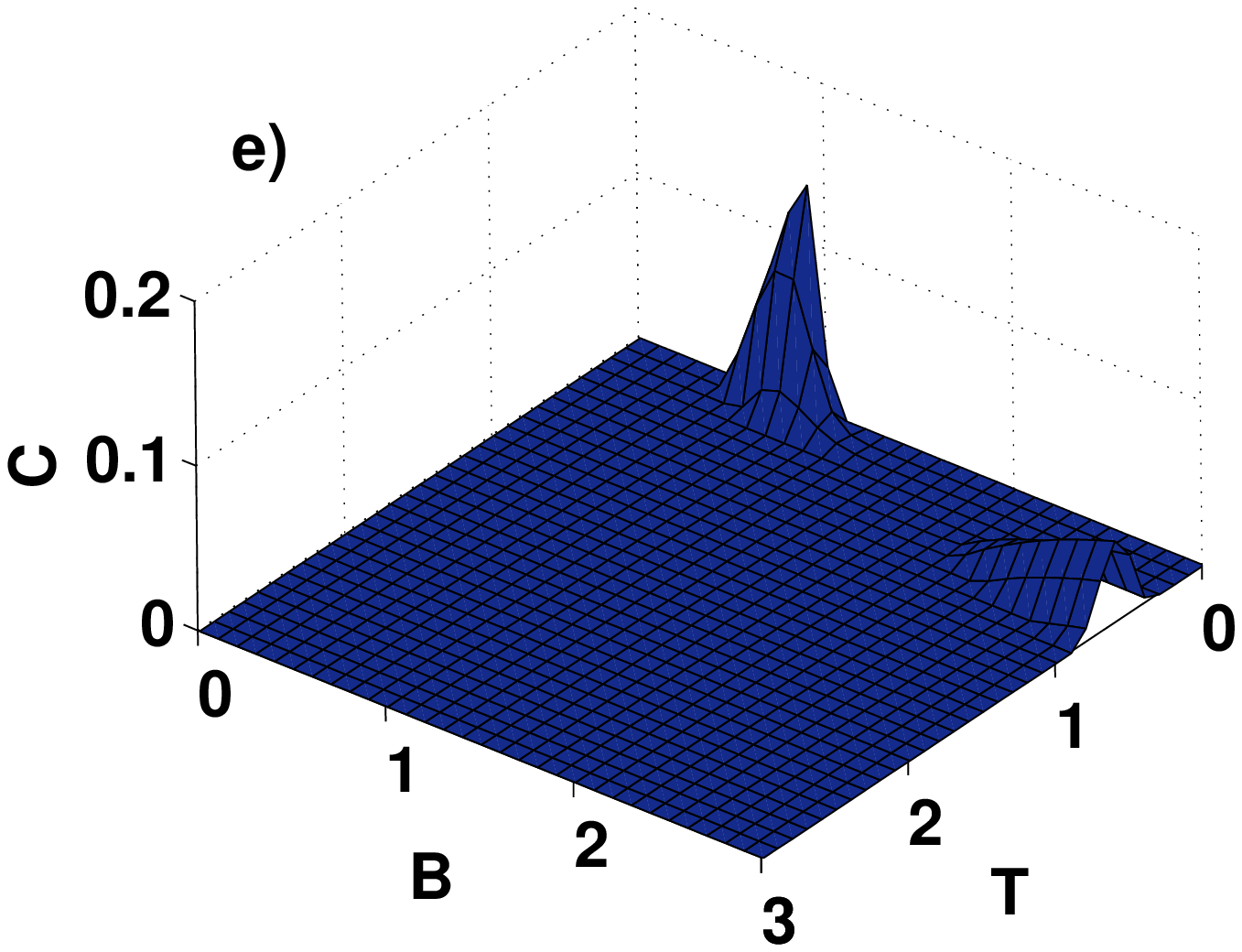,width=0.5\columnwidth,clip=} \\
\epsfig{file=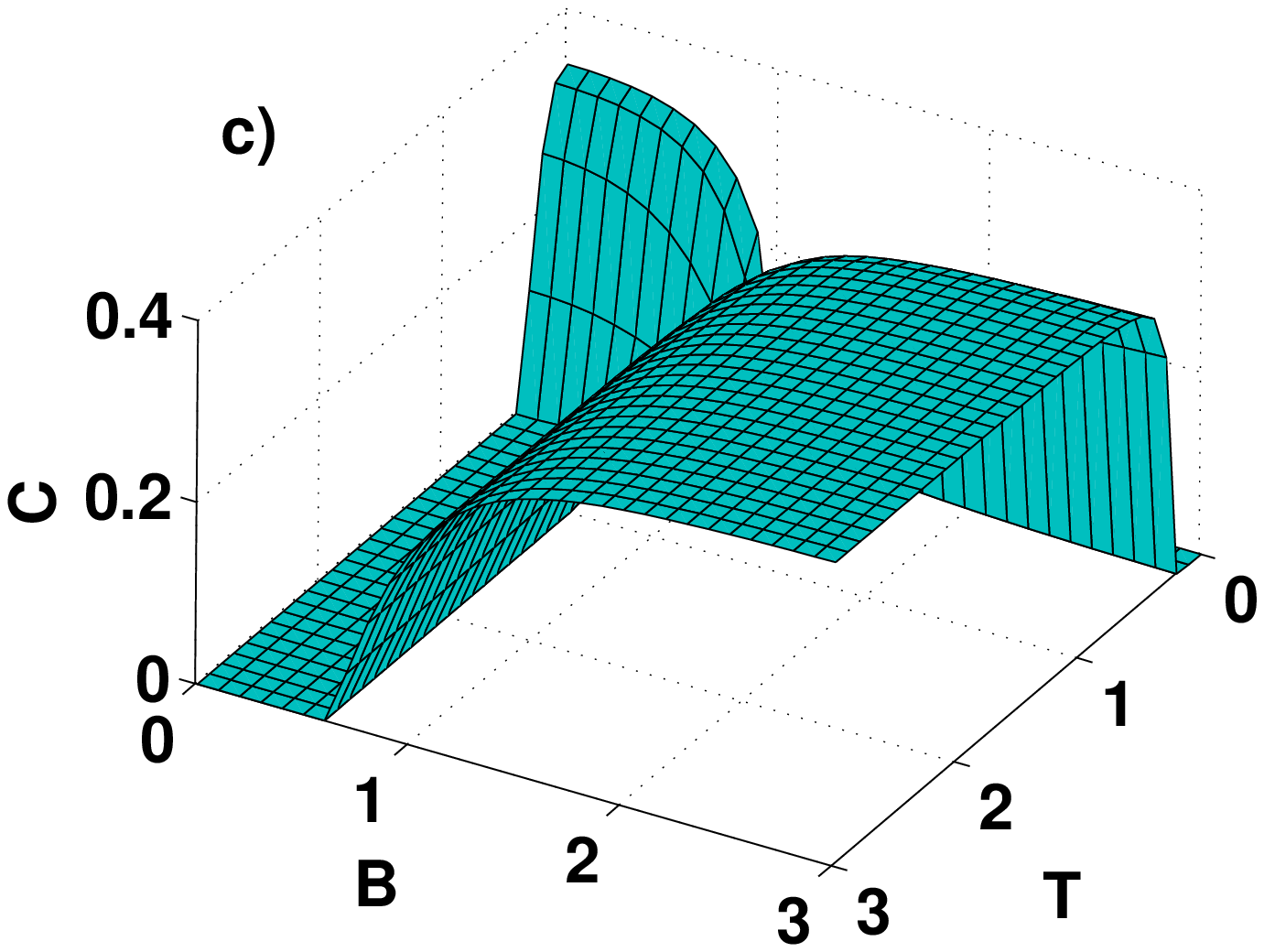,width=0.5\columnwidth,clip=} &
\epsfig{file=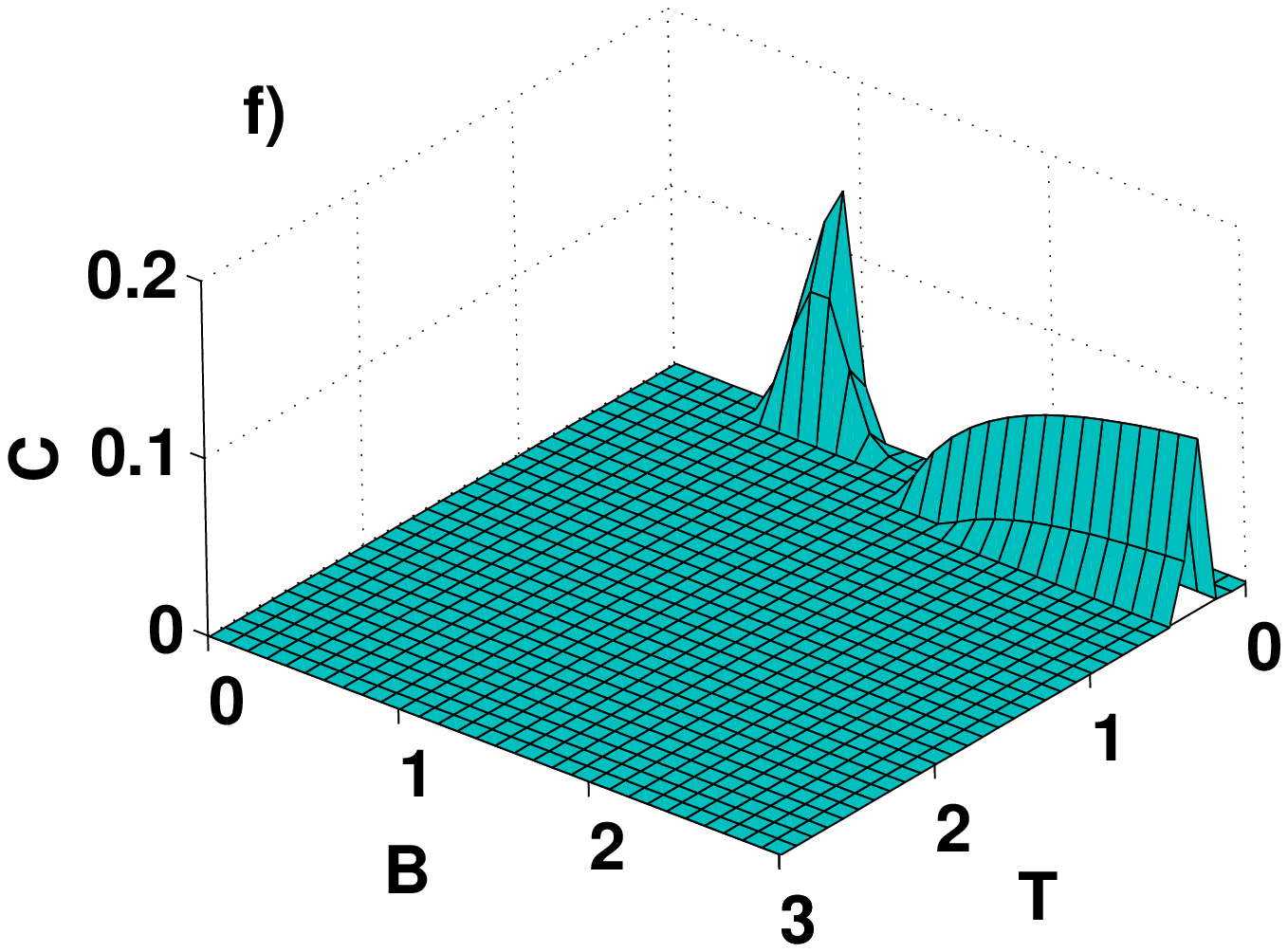,width=0.5\columnwidth,clip=} \\
\end{tabular}
\caption{Nearest and next nearest neighbour
concurrence respectively with a) and d) when $\gamma=0$, b) and e)
when $\gamma=1$, and c) and f) when $\gamma=2$. We have set $J=1$ throughout.}
\label{Fig:SSConc}
\end{figure}

We plot the concurrence against the magnetic field and steady state
temperature in Fig. \ref{Fig:SSConc}. We find for nearest neighbour
concurrence, Figs. \ref{Fig:SSConc} a), b) and c), that the amount
of entanglement decreases at low $T$ and $B$ with increasing
$\gamma$, and also that the region of entanglement decreases. The
amount and region of entanglement increases with $\gamma$ when $T$
and $B$ are higher. The plot coincides with both witnesses which
detect entanglement at high $B$ and $T$. Fig. \ref{Fig:EntWitNormal}
shows that the witness also detects a region of entanglement at low
$B$ and $T$ that decreases with increasing $\gamma$ which is reflected in
Fig. \ref{Fig:SSConc}. Fig.
\ref{Fig:EntWitSS} however, also detects a region of entanglement at
low magnetic field, and higher $T$ and $\gamma$, which indicates
this witness detects entanglement which cannot be categorised as
nearest neighbour. Therefore we consider next nearest neighbour
concurrence plotted in Fig. \ref{Fig:SSConc} d), e) and f). Both the amount and
region of entanglement are reduced compared to nearest neighbour
concurrence. The peak at low temperature remains constant for each
value of $\gamma$, while as $\gamma$ is increased, new regions of
entanglement appear at low temperature and higher magnetic field.
Thus for next nearest neighbour entanglement, the introduction of an
energy current again increases the amount of entanglement. However,
this does not explain the region of entanglement at low $B$ and
higher $T$ and $\gamma$ detected in Fig. \ref{Fig:EntWitSS}. It
would be interesting to determine the type of entanglement here.


We have introduced a nonequilibrium steady state entanglement
witness and using an $XX$ spin chain, have demonstrated how such
a witness works. We have shown that introducing an energy
current increases the entanglement detected in this system
at high steady state temperature, and
increases entanglement itself on consideration of the concurrence.

\emph{Acknowledgements:} We thank Marie Curie and the JSPS
for financial support, and K. H\"{a}rk\"{o}nen, B. Garraway,
W. Lange and M. Murao for useful discussions.


\begin{thebibliography}{99}

\bibitem{manybody}
L. Amico et al.
Rev. Mod. Phys. \textbf{80}, 517 (2008)

\bibitem{Gemmer}
J. Gemmer, M. Michel and G. Mahler,
\emph{Quantum Thermodynamics} (Springer-Verlag, Berlin, 2004)

\bibitem{ent_wit}
G. Toth,
Phys. Rev. A \textbf{71}, 010301(R) (2005)

\bibitem{vlat}
\v{C}. Brukner and V. Vedral,
Arxiv: quant-ph/0406040

\bibitem{Wiesniak}
M. Wie\'{s}niak, V. Vedral and \v{C}. Brukner,
New J. Phys. \textbf{7}, 258 (2005)

\bibitem{HideJ} J. Hide and V. Vedral,
Phys. Rev. A \textbf{81}, 062303 (2010)

\bibitem{Oono} Y. Oono and M. Paniconi,
Prog. Theor. Phys. Suppl. \textbf{130}, 29 (1998)

\bibitem{Sasa} S. Sasa and H. Tasaki,
J. Stat. Phys. \textbf{125}, 125, (2006)

\bibitem{Hardy}
R. Hardy,
Phys. Rev. \textbf{132}, 168 (1963)

\bibitem{Saito}
K. Saito, S. Takesue and S. Miyashita,
Phys. Rev. E \textbf{54}, 2404 (1996)

\bibitem{Michel}
M. Michel, J. Gemmer and G. Mahler,
Int. J. Mod. Phys. B, \textbf{20}, 4855 (2006)

\bibitem{WuAndSegal}
L. Wu and D. Segal,
J. Phys. A, \textbf{42}, 025302 (2009)

\bibitem{Kosov}
D. S. Kosov,
J. Chem. Phys. \textbf{120}, 7165 (2004)

\bibitem{Antal1}
T. Antal, Z. R\'acz and L. Sasv\'ari,
Phys. Rev. Lett. \textbf{78}, 167 (1997)

\bibitem{Antal2}
T. Antal, Z. R\'acz, A. R\'akos and G. M. Sch\"utz,
Phys. Rev. E \textbf{57}, 5184 (1998)

\bibitem{Eisler}
V. Eisler and Z. Zimbor\'as,
Phys. Rev. A \textbf{71}, 042318 (2005)

\bibitem{ZWang}
Z. Wang et al.
Physica A \textbf{387}, 2197 (2008)

\bibitem{Casas}
J. Casas-Vazquez and D. Jou,
Braz. J. Phys. \textbf{27}, 547 (1997)

\bibitem{Kita}
T. Kita,
J. Phys. Soc. Jpn. \textbf{71} 1795 (2002)

\bibitem{Katsura}
S. Katsura,
Phys. Rev. \textbf{127}, 1508 (1962)

\bibitem{expQ}
A. V. Sologubenko et al.
Phys. Rev. B \textbf{62}, R6108 (2000)

\bibitem{OConnor}
K. M. O'Connor and W. K. Wootters,
Phys. Rev. A \textbf{63}, 052302 (2001)

\bibitem{Barouch}
E. Barouch and B. M. McCoy,
Phys. Rev. A \textbf{3}, 786 (1971)



\end{thebibliography}
\end{document}